\documentclass[preprint,pra,superscriptaddress]{revtex4}

\usepackage[dvips]{graphicx} 
\usepackage{amsfonts}
\usepackage{amssymb}
\usepackage{amscd}
\usepackage{amsmath}    
\usepackage{enumerate}
\usepackage{epsfig}
\usepackage{subfigure}

\begin{document}

\title{Improved Data Post-Processing in Quantum Key Distribution and Application to Loss Thresholds in Device Independent QKD}

\author{Xiongfeng Ma}
\email{xfma@iqc.ca}
\author{Norbert L\"utkenhaus}
\email{nlutkenhaus@uwaterloo.ca}
\affiliation{%
Institute for Quantum Computing and Department of Physics and Astronomy, \\
University of Waterloo, 200 University Ave W., Waterloo, ON, Canada N2L 3G1 \\
}%

\begin{abstract}
Security proofs of quantum key distribution (QKD) often require post-processing schemes to simplify the data structure, and hence the security proof. We show a generic method to improve resulting  secure key rates by partially reversing the simplifying post-processing for error correction purposes. We apply our method to the security analysis of device-independent QKD schemes and of detection-device-independent QKD schemes, where in both cases one is typically required  to assign  binary values even to lost signals. In the device-independent case, the loss tolerance  threshold is cut down by our method from 92.4\% to 90.9\%.The lowest tolerable transmittance of the detector-device-independent scheme can be improved from 78.0\% to 65.9\%.

\end{abstract}

\maketitle

\section{Introduction}
Quantum key distribution (QKD) \cite{BB_84,Ekert_91} provides a means of distributing secure keys between two distant parties, Alice and Bob. It uses a quantum channel and an
authenticated classical channel. The security of specific abstract QKD protocols has been proven in literature under varying definitions of security \cite{Mayers_01,LoChauQKD_99,ShorPreskill_00}. A clear framework for security proofs including finite-size effects has been put forward by Renner \cite{Renner_Thesis_05}.  For a review of the subject, refer to \cite{GRTZ_02,LoLut_review_07,RevQKD_Lut_08} and references therein.

When it comes to real optical implementations, security proofs have to be able to take non-ideal devices  into consideration. A lot of effort has been made to achieve the security of QKD with realistic devices \cite{KoashiPreskill_03,ILM_07,GLLP_04,TT_Thres_08,BML_Squash_08}. One  approach is based on exact models of the devices. As security proofs have to apply to the infinite-dimensional Hilbert space of optical modes, one seeks methods to simplify the analysis. One method is to apply coarse-graining of data, by which we mean a postprocessing of data. Let us give two examples for this coarse-graining.

The first example is the random assignment of double-click events in the BB84 protocol with threshold detectors. This assignment allows the construction of a squashing model \cite{BML_Squash_08} so that the security of the optical implementation can be tied directly to a corresponding abstract qubit protocol.

The other example is the treatment of loss in QKD. The possibility of transmission and detection losses can be easily incorporated in the security proofs as long as it holds that the loss probability is independent of the chosen measurement basis. However, as demonstrated in \cite{MAS_Eff_06,QFLM_TimeShift_07,Timeshift_Exp_08,Makarov:Bright:09} this assumption is violated in typical implementations by adversarial action. As a result, QKD schemes can be completely broken if they do not address this point. One way to patch this weakness is to assign measurement outcomes even to lost signals. The typical choice is a random assignment. In this way, signal loss  is converted into an effective error rate on the data. Using this patch, it is now possible to circumvent the need for precise characterization of devices in the  paradigm of device-independent (originally known as self-testing) QKD schemes. Since the early work by Mayers and Yao in 1998 \cite{MayersYao_98}, a few more realistic \cite{AGM_Bell_06,Acin:DeviceIn:07} and generalized \cite{McKague_Selftest_10} schemes have been proposed. Recently, the security analysis of device-independent schemes has  made progress \cite{Acin:DeviceIn:07,Pironio:DeviceIn:09}. Besides a missing unconditional security proof, the main drawback of these schemes are their severe constraints on physical devices in practice. For example, the security analysis of Pironio et al. \cite{Pironio:DeviceIn:09} gives a tolerable error rate is $7.1\%$ which translates to a  required minimum transmittance of $92.4\%$. Less restrictive schemes are detection-device-independent schemes which assume some generic structure on the source side, but makes no assumption on the detection devices. Examples of this approach include \cite{Mayers_01,Koashi:Comp:09,berta10a}. The scheme by Koashi tolerates an error rate of $11\%$ and a lower transmittance of $65.9\%$.

The postprocessings mentioned above (random assignment of data to double-clicks, random assignment of lost signals) are done for convenience. So for the purpose of the security proofs, we erase the history of the assignment, meaning that we ignore the knowledge which of the data have been affected by the postprocessing.  It is clear that security proofs can be also obtained without this erasure, however, these proofs will be more complicated to obtain. In this paper we address the question whether we can improve the key rate by making use of our knowledge of the positions within the data string that, for example, have been assigned random values. We will demonstrate a generic way of making use of the knowledge, without the need to revisit the full security proof. Of course, this approach will work only if the QKD protocol and the security proof follows some generic structure. In Section \ref{Sec:Det:Proofs} we outline these generic structures. Then, in Section \ref{Det:Sec:Advpp} we provide the generic method to use our extra knowledge to improve the key rate. We then illustrate the effect of our method for two examples, for the detector-device-independent scheme in Sec. \ref{Sec:Det:DDI}, and for the full device-independent scheme in Sec. \ref{Sec:Det:FDIS}.

\section{Protocols and Security proofs} \label{Sec:Det:Proofs}
QKD protocols typcially involve two phases: the quantum phase, in which quantum signals are distributed and measured, resulting in correlated classical data shared between Alice and Bob, and a classical phase, where these classical data are processed by classical communication protocols. The classial phase has the goal to establish a secure key. By definition, a secure key should be \emph{identical} between Alice and Bob, and \emph{private} (unknown  to an eavesdropper, Eve). For an example of data processing procedures,  refer to \cite{Renner_Thesis_05,Finite:Short:09,Finite:Long:10}. A typical classical data processing can be divided into three steps:
\begin{enumerate}
\item
{\bf data pre-processing}, which includes
\begin{enumerate}
\item {\bf Sifting processes:} any post-selection of signals, for example  based on basis settings or non-detection events. Sifting always uses two-way communication.
\item {\bf Coarse graining:} data processing aiming at simplifying the data structure, such as random bit assignments for double clicks, or random bit assignments for non-detected signals; coarse graining is always a local process and does not involve any communication. All following steps are based on the coarse-grained data only,
\item {\bf advanced pre-processing:}  further pre-processing, locally or by two-way communication, based on the coarse-grained data, for example advantage distillation \cite{maurer93a}, especially the so-called {\em B steps} \cite{TwoWay_03,TwoWay_06}),  and adding noise \cite{KGR_noise_05};
\item {\bf parameter estimation:}parameter estimation of the correlations shared between Alice and Bob, thus drawing limits on the correlations between Alice and Eve by quantum mechanics. This includes the estimation of the error rate, but also includes the decoy state analysis \cite{Hwang_03,Decoy_05,Wang_05}.
\end{enumerate}

\item
{\bf error correction}, which ensures the key shared by Alice and Bob to be \emph{identical};

\item
{\bf privacy amplification}, which ensures the key to be \emph{private}.
\end{enumerate}

In this work, we assume that the step of error correction is performed in a uni-directional way from one party to the other, without loss of generality from Alice to Bob. As a result, the final key can be determined by Alice already after the data pre-processing step, as she sends out error correction information to Bob which enables Bob to correct his data to Alice's, and the privacy amplification step can be done with a hashing function chosen by Alice.   Our results may be extended to  scenarios without this assumption, as long as Alice's data determine the final key.

There are many security proofs for QKD protocols that follow the outlined procedure, see references in \cite{GRTZ_02,LoLut_review_07,RevQKD_Lut_08}. These proofs are using different techniques and give a rate $R$ at which secure key can be generated. We  illustrate our finding in the infinite key limit, denoted as  rate $R_\infty$, although our method will be directly applicable also for any analysis that includes finite size statistics. For our generic protocol, this key rate takes the form
\begin{equation} \label{Det:Security:R}
R_\infty \geq H(A)-f H(A|B)-I_{pa},
\end{equation}
where $H(A)$ entropy of Alice's data after the data pre-processing step. Further,  $H(A|B)$ is the entropy on Alice's data given Bob's data, so that this term amounts to the minimum number of bits (Shannon limit) that Alice has to send to Bob per retained signal in order for Bob to be able to correct his data to match it to Alice's data. The factor $f \geq 1$ is an efficiency factor that characterizes the actual error correction protocol that has been followed. The Shanon limit corresponds to $f=1$, while in practice we find typical values of $f \in (1.1, 1.25)$.  The last term,  $I_{pa}$, is some measure of Eve's information on the key, leading to a key reduction in privacy amplification. The form of $I_{pa}$ depends on the exact form of the security definition of the key, the exact QKD protocol, and potentially also on the security proof technique. It is  the latter reason that we refer to these key rates as lower bounds, as any valid security proof guarantees that at least this amount of secret key can be extracted during the QKD protocol, while improved security proofs may give higher secret key rates without changing the protocol.

 There are three widely used approaches for security analysis:
\begin{enumerate}
\item {\bf Entanglement distillation based:} Shor-Preskill \cite{ShorPreskill_00}, based on \cite{BDSW_96,LoChauQKD_99}, with $I_{pa}=h[e_p]$, where $e_p$ is the phase error rate of qubit signals and $h[x]=-x\log_2(x)-(1-x)\log_2(1-x)$ is the binary entropy function; This approach is designed for fully characterized devices and typically requires specifically constructed error correction methods (e.g. linear codes).

\item 
{\bf Entropy based: characterized devices} Based on ideas by Ben-Or, this approach has been advanced by Devetak and Winter \cite{DevetakWinter:05}, and made rigerous by Renner \cite{Renner_Thesis_05}. In the infinite key limit, we obtain  with $I_{pa}=\chi_E$ and $\chi_E$ is the Holevo bound \cite{Holevo:Bound:1973} on Eve's information about Alice's key; this approach has been also used to prove device-independent QKD protocols to be secure against collective attacks \cite{Pironio:DeviceIn:09}.

\item {\bf Entropy based: Complementarity} This approach has been first proposed by Koashi \cite{KoashiPreskill_03,Koashi:Comp:09} and recently put into a rigorous framework by Tomamichel et al. \cite{tomamichel11suba}. We obtain for qubits again $I_{pa}=h[e_p]$. This approach has been suggested to remain  valid even if uncharacterized detection devices are used.
\end{enumerate}

In all these proofs, the key rate can be viewed as the rate at which signals emerge from the initial data pre-processing, shortened by the amount of information of error correction sent from Alice to Bob and also shortened by a privacy amplification term $I_{pa}$. As stated before, for our analysis the exact details of the security analysis leading to the term $I_{pa}$ will not be important, as we will deal with the error correction part of the protocol. For this, however, it is important to understand how the effective key rate $R_\infty$ comes about. Let us outline some of the approaches that deal with the influence of error correction on the key rate.

The first method uses encryption of the data sent from Alice and Bob by a one-time pad during error correction. This idea has been proposed in \cite{Lutkenhaus:practical:99} and later refined by \cite{lo_decouple_2003}. This approach allows to decople error correction from privacy amplification. Then privacy amplification needs only to address the initially available correlations between Alice and Eve. As we use one-way error correction, Alice's data are not affected by this error correction at all. So we generate a secret key at rate $H(A)-I_{pa}$, but we use up secret key at a rate of $f H(A|B)$ to encrypt the error correction information, resulting in the net rate shown in Eqn.~(\ref{Det:Security:R}).

The second method is to keep track of the amount of information that becomes available to Eve during error correction. In this approach, privacy amplification has to shorten the key in order to cut out not only Eve's initial correlations with Alice's data, but also those correlations between Alice and  Eve that result from Eve listening to the extra information becoming available during error correction. The result is again the net rate Eqn.~(\ref{Det:Security:R}), though the argumentation behind this form differs between the different security approaches. In the entanglement based approach, the argumentation follows directly from the structure of quantum error correction codes that are used to effectively distill entanglement. In the two entropy based approaches, instead, one uses results that show that the entropy of Eve's system conditioned on Alice's data can be reduced at most by one bit per bit of data announced by Alice during error correction. As the amount of required key shortening during privacy amplification depends exactly on Eve's entropy about the key before privacy amplification, we obtain again the key rate shown in Eqn.(\ref{Det:Security:R}).

\section{Advanced Efficient Error Correction} \label{Det:Sec:Advpp}
As we have seen in the previous discussion, the key rates that result form the various security proofs depend  only on the actual amount of error correction information that is being sent from Alice to Bob. Our key observation is that this amount of information can be reduced if Bob accesses his refined knowledge about his detection events. Moreover, accessing the refined information by Bob for the purpose of error correction in this setting does not affect the correlations between Alice and Eve, and hence does not affect the term $I_{pa}$. We will now make these statements more precise. For our arguments, we continue to refer for illustrative purposes to the asymptotic key rate in QKD, but we emphasize again that all arguments also hold in the case of finite key sizes.

We start with the cost of error correction. We denote by $B^{(c)}$ the refined data known to Bob  after sifting, coarse graining and advance pre-processing. Then $B^{(r)}$ are the same data, extended by the refined data on which the coarse-graining was based.  Then it follows from the data-processing inequality \cite{Cover2006} that
 $$H(A|B^{(c)})\geq H(A|B^{(r)})$$
so that we have the opportunity to  enable Bob to reconcile his data with Alice's data string with less error correction data by making used of the refined knowledge ob Bob.

\subsection{Background}
Next, we need to investigate the effect the use of the refined data have on the full security proof, and thus on the term $I_{pa}$. If we use the method of encrypting the error correction information using the one-time pad encryption of the error correction information, then it is clear that if the larger amount $H(A|B^{(c)})$ of information is sent encypted and the resulting key is secure, then the security is not affected if the smaller amount of information according to $H(A|B^{(r)})$ is sent, as from Eve's point of view nothing changes. However, as a result, the net key rate increases. Note that also the amount of encrypted error correction information does not change Eve's view: Eve can predict the statistics of the refined information on Bob's side from her eavesdropping action, and also Alice can only make use of this statistics to design the error correction information.

If we use an entanglement based approach, including a structured quantum error correction method, then  the correction of bit-errors decouples from the correction of phase-errors. The refined data then serve as a side-information to Bob to help him to be more efficient in correcting the bit-errors. The correction of phase-errors (and hence the privacy amplification) is not affected by this.

Finally, using entropy-based security proof approaches, note that again the refined knowledge can be considered as local side-information at Bob's side. As it is not known to Eve or even Alice, it does not affect the initial correlations between Alice's and Eve's data. The entropy-based security proofs only count the actual number of error correction bits that Alice sends out to Bob to allow Bob to perform error correction. If this number goes down, the effective secret key rate goes up by the same amount.

Note that all these arguments also hold in the case of finite key sizes as in all security proof approaches the final key size depends on the actual number of bits exchanged during error correction. Moreover, the decoupling arguments for error correction information and $I_{pa}$ holds also in both cases. Therefor our method is compatible with finite size security proofs including \cite{Hayashi:Finite:06,Scarani:Finite:08,Finite:Short:09,Finite:Long:10,Renner_Thesis_05,tomamichel11suba}.

\subsection{BB84 protocol } \label{Sec:Det:BB84}
To illustrate the effect of our method, let us start by considering the BB84 protocol \cite{bennett84a}. The secure key rate for the qubit based protocol is given by
\begin{equation}
\label{Eq:BB84standard}
R_{BB84}^{qubit} = 1 - h[e] - h[e]
\end{equation}
where $e$ is the observed quantum bit error rate. This rate can be proven to be secure by many different methods \cite{Mayers_01,ShorPreskill_00,Koashi:Comp:09,Renner_Thesis_05}, where the results of Mayers and Renner can also be shown to hold for lossy channels. In the latter case, the base of the secret key rate are the detected signals. We separated the individual terms corresponding to error correction at the Shannon limit as  $H(A|B)= h[e]$ from the privacy amplification term $I_{pa}= h[e]$.

In optical systems, even using ideal single photon sources as qubit sources, we need to deal with the fact that the detectors operate on optical modes. By assigning  double clicks \cite{Lutkenhaus_99DoubleClick} to randomly assigned binary values, one can use the existence of a squashing model \cite{BML_Squash_08,TT_Thres_08} for the photo-detector set-up to uplift the secure key rate for the qubit protocol over lossy qubit channels to the the same protocol over lossy optical channels with threshold detectors.

Doing the coarse-graining of assigning double-clicks to single click events, we find an effective quantum bit error rate for the coarse grained data as
\begin{equation} \label{Det:Pre:QBER}
\begin{aligned}
e^{(c)} &= P _s e_s + (1-P_s) \frac{1}{2},
\end{aligned}
\end{equation}
where $P_s$ is the rate of single clicks within the detected signals, which exhibit some error rate $e_s$ so that $1-P_s$ is the rate double clicks, which lead through  random-bit assignment to an error rate of $1/2$. This results in a key rate
\begin{equation}
\label{Eq:CoarseBB84}
R_{BB84}^{(c)}= 1 - 2 h[e^{(c)}] \; .
\end{equation}

We now give Bob access to refined data, so that he knows at which positions he assigned random single click outcomes instead of the double click outcomes. The entropy $H(A|B^{(r)})$ corresponds to that of an erasure channel with error rate $e_s$, so we find
\begin{equation}
\label{eqn:BB84:refined}
H(A|B^{(r)}) = P_s h[e_s] + (1-P_s) \;.
\end{equation}
With that, the key rate improves over Eq. (\ref{Eq:BB84standard}) to
\begin{equation}
R_{BB84}^{(r)} = P_s \left( 1-h[e_s] \right) - h[e^{(c)}] \; .
\end{equation}
As discussed before, the data processing inequality guarantees that this method improves the secret key rate. The number of double-clicks in typical experiments is not very high, so the difference between the two secure key rates is not very big. Our next example will show a scenario where the difference is more important, as there we will be forced to deal with loss of signal. Signal loss is by far the dominating effect in QKD secure key rates compared to the effect of errors within the detected signals.

\subsection{Detection-device-independent scheme} \label{Sec:Det:DDI}
In many QKD security proofs, one starts with a full characterization of sources and detection devices. This scenario can be relaxed by having only characterized sources, while the detection devices remain uncharacterized. A security proofs following these lines is the one by Mayers \cite{Mayers_01}, but also the proofs of  Koashi \cite{Koashi:Comp:09} and Tomamichel \cite{tomamichel11suba} have been suggested to have this property. However, in order to deal with transmission and detection lossses, one has either to assume that the detection efficiency of the receiver is basis independent,  and then can discard all lost signals,  or one does not adopt any additional assumption, and then has to coarse-grain the non-detection event into the (binary) outcome events in order for these proofs to apply directly.

We are interested in the case where we do not make additional assumption on the receiver, and instead  randomly assign binary values  to no-click and double click events. Koashi has indicated that, given a source which emits signals such that the density matrix averaged over the signals of each basis are independent of this basis, the secret key rate for the standard coarse-grained  data processing  is given by once again by Eq.~(\ref{Eq:CoarseBB84}) with the corresponding error rate  given by Eq.~(\ref{Det:Pre:QBER}). In that error rate, the single click probability $P_s$ and error rate $e_s$ is now the complement not only of the double clicks, but also of the no-click events, all of which are now coarse-grained by mapping  them  into random events.

To illustrate the improvement we perform a simulation of the observed parameters to predict the key rates. For this purpose we will neglect detector dark counts, as our simulations will give secret keys only for total transmissivities of 50\% and more and therefore the rate of detected events will, typically,  be many orders of magnitude higher than the dark count rate. The key rates are shown in the infinite-key limit with error correction being performed in the ideal case reaching the Shannon limit. For sources we assume perfect single-photon sources, so that $P_s = \eta$ where $\eta$ is the single-photon transmissivity of the overall system, including transmission and detection loss. The result also hold for parametric-downconversion sources as long as the heralding set-up assures that for the heralded signals the average density matrix for the two basis is still basis independent. In this picture, the single-event error rate $e_s$ is due to factors such as misalignment and decoherence mechanisms in the channel.

\begin{figure}[hbt]
\centering \resizebox{10cm}{!}{\includegraphics{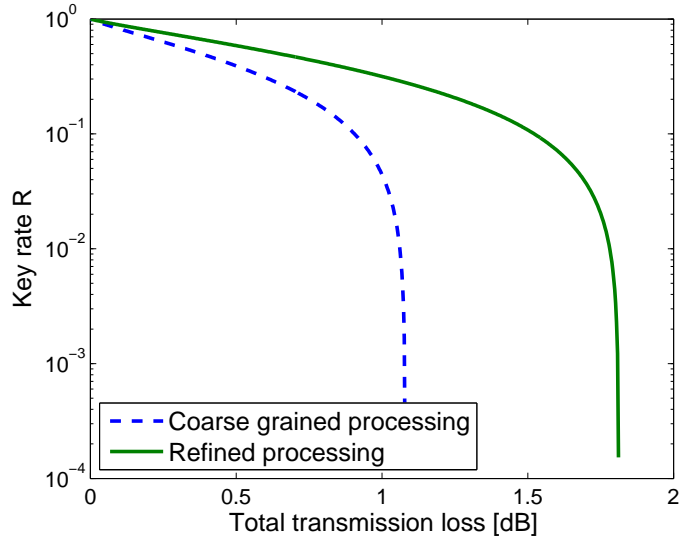}} \caption{Plot of key rate of the two data processing schemes for the detection-device-independent scheme. The lowest tolerable transmittance of the coarse-grain processing scheme is 78.4\%, while that of refined-grain processing scheme is 65.9\%. Here we assume there is no error for single-clicks ($e_s=0$) and the error correction reaches the Shannon limit.} \label{Fig:Eff:Reta}
\end{figure}

 The simulation result is shown in Figure \ref{Fig:Eff:Reta} for the ideal case where we have no errors at all that orginate from single clicks ($e_s=0$). Of particular interest  is the threshold for the transmittance above which secret keys can be generated. For the standard coarse-graining data processing one finds for this threshold the value of 78.4\%  (corresponding to about 11\% error rate), which is consistent with the result of
Shor-Preskill's proof \cite{ShorPreskill_00}. The transmittance threshold for the
advanced data processing scheme based on the refined data is case is 65.9\%.  From the time-shift attack, we know that the lower bound of tolerable transmittance for this untrusted detection device case is 50\% \cite{EffLoop_08}.

\begin{figure}[hbt]
\centering \resizebox{10cm}{!}{\includegraphics{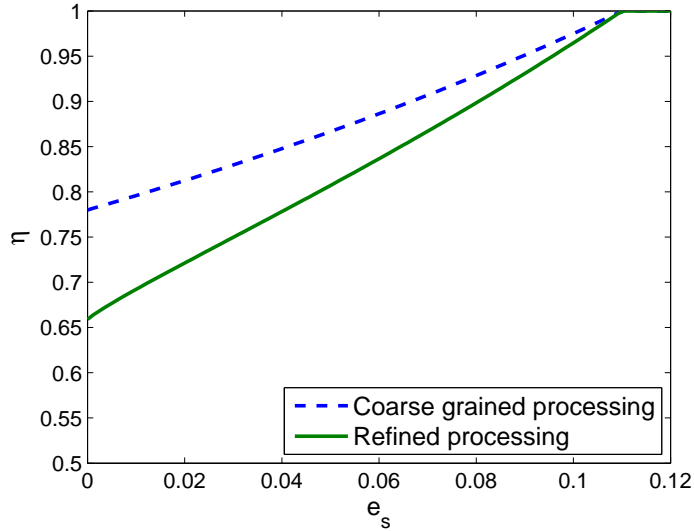}} \caption{Plot of the tolerable $e_s$ and $\eta$ for the two data processing schemes.} \label{Fig:Eff:Lower}
\end{figure}

To show the effect of a non-zero error rate within the single-click events, we plot the thresholds for the two methods as a function of the single-click event error rate $e_s$ in Figure~\ref{Fig:Eff:Lower}. There is no positive key for $e_s>11.0\%$, which is consistent with the result in  \cite{ShorPreskill_00}.

\subsection{Fully device-independent scheme}
\label{Sec:Det:FDIS}
In fully device-independent QKD, none of the devices of sender and receiver are characterized. These schemes are entanglement based QKD schemes, so that Alice and Bob have two detection devices each, while the adversary Eve is in control of the source. Alice and Bob  each choose actively between their uncharacterized devices.  Security proofs for these schemes, assuming collective eavesdropping attacks, have been given in \cite{Acin:DeviceIn:07,Pironio:DeviceIn:09}. This proof assumes that the data are coarse-grained into binary outcomes. In constrast to earlier coarse-graining, we will in this case follow reference  \cite{Pironio:DeviceIn:09} and assign a pre-agreed {\em fixed} binary value to lost signals, rather than random binary values, as this gives a slight advantage. Note that this choice also influences the Shannon entropy $H(A)$ in Eqn. (\ref{Det:Security:R}), which is now less than one.  The privacy amplification term is given by
\begin{equation} \label{Det:Adv:diR}
\begin{aligned}
I_{pa}=h\left[\frac{1+\sqrt{(S/2)^2-1}}{2}\right]
\end{aligned}
\end{equation}
where $S$ is the CHSH \cite{CHSH.Bell.69} Bell parameter. To simulate this parameter for experiments, we neglect the double click events, assuming a perfect photon-pair source. Within single clicks events on both sides, there is an observed error rate $e_s$ when measuring in the same basis. We assume this error rate to be the parameter of a depolarizing channel in order to predict the correlations when Alice and Bob do measure in different bases in order to determine $S$. The depolarizing channel maintains the signal perfectly with probability $1- 2  e_s$, while it randomizes the signal with probability $2 e_s$. A perfect signal leads to a Bell parameter $S=2\sqrt{2}$, while a randomized signals gives $S=0$.  The tranmittance between source and Alice and Bob is given by $\eta_A$ and $\eta_B$ respectively.

There are three contributions to the value of $S$: The first describes that case when both sides detect single clicks, which happens with probability $P_s= \eta_A \; \eta_B$. The $S$ parameter in this case is given by $2 \sqrt{2} (1-2 e_s)$. The second contribution comes happens if both signals are lost, which happens with probability $(1-\eta_A)(1-\eta_B)$. Since the binary outcomes are predetermined in this case and perfectly correlated, we find $S=2$. Finally, the third contribution has one detected photon and one lost photon. The fixed assignement leads to uncorrelated data between Alice and Bob, and thus to $S=0$. Overall, we find then for the Bell parameter
\begin{equation}
S = 2\sqrt2(1-2e_s)\eta_A \eta_B+2(1-\eta_A) (1-\eta_B) \; .
\end{equation}

The key is generated by measurements where Alice and Bob perform measurements in identical bases. Therefore the error rate of the coarse-grained data is given by
\begin{equation}
e^{(c)} = P_s e_s + \left((1-\eta_B)\eta_A +(1-\eta_A)\eta_B \right) \frac{1}{2} \;.
\end{equation}
Note that positions where neither Alice nor Bob do detect a photon do no contribute to an error rate, as Alice and Bob would assigned the same bit value to these events. However, the entropy of Alice's data is now given by $H(A) = h\left[\frac{1}{2} \eta_A + (1-\eta_A)\right]$.

Overall, this particular  coarse grained data processing leads to the key rate
\begin{equation}
R_{DI}^{(c)} = h\left[\frac{1}{2} \eta_A + (1-\eta_A)\right] - h[e^{(c)}] -h\left[\frac{1+\sqrt{(S/2)^2-1}}{2}\right]\; .
\end{equation}

\begin{figure}[hbt]
\centering \resizebox{10cm}{!}{\includegraphics{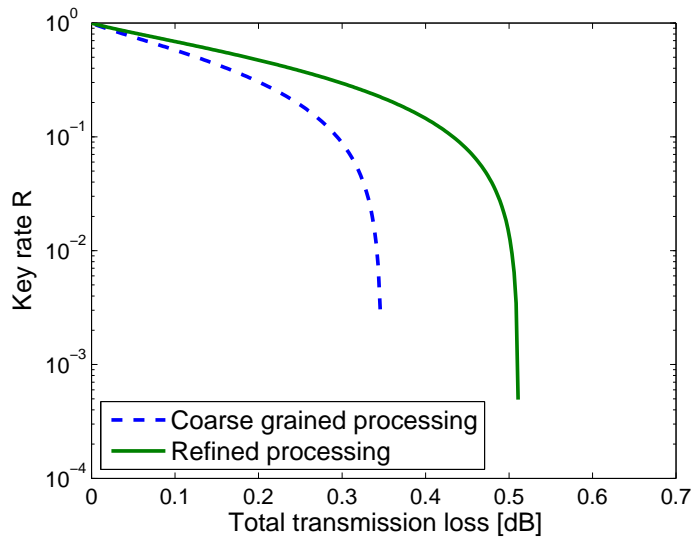}} \caption{Plot of key rate for the device-independent QKD scheme with a double link setup and $\eta_A=\eta_B=\eta$. The lowest tolerable transmittance of a single link for  the coarse-grain processing scheme is 92.4\%, while that of refined-grain processing scheme is 90.9\%. Here we assume there is no error for single-clicks ($e_s=0$) and the error correction reaches the Shannon limit.} \label{Fig:Eff:DiRdB}
\end{figure}

For the refined data processing, Alice keeps her coarse graining method as these data define the key. However, Bob now accesses the refined data. The key rate  is shown in Figure \ref{Fig:Eff:DiRdB}, using the same modelling as in the previous section, including the choice of $e_s=0$. In contrast to the detection device indepent case, we now assume the source to be symmetric located with respect to Alice and Bob, leading to the choice $\eta_A = \eta_B \equiv \eta$. The tolerable transmittance$\eta$ for a single link for the standard processing is 92.4\% (corresponding to a total transmittance of $\eta^2$, or 82.6\%), which is consistent with the result shown in \cite{Acin:DeviceIn:07}. With the advanced data processing scheme, the tolerable single link transmittance can be improved to 90.9\%.

\section{Conclusion} \label{Sec:Eff:Conclusion}
We presented a generic method to improve the secret key rate for quantum key distribution systems. This method applies in situations, where information of the receiving party has been coarse-grained in order to apply some security proof method. We showed that the refined information can be accessed for error correction purposes, thus increasing the effective key rate.

While we demonstrated our method in the scenario of one-direction error correction, we would like to point out that the framework of Renner \cite{Renner_Thesis_05} has as its essential feature only that the final key is derived from the data on one side. This can also be achieved by two-way error correction mechanism and the key rate depends only on the actual amount of information leaked to the adversary during error correction. Therefore, our method can be extended to this situation even without encryption of the error correction information. For the same reason, the extension to entanglement based protocols is contained in our analysis.

\subsection*{Acknowledgements}
We thank Hoi-Kwong Lo for enlightening discussions. This work is supported by NSERC via the Innovation Platform Quantum Works, the  Discovery grant programme, and the Strategic Project Grant FREQUENCY, and by the Ontario Research Fund (ORF).

\bibliographystyle{apsrev}



\end{document}